\documentclass[aps,twocolumn,showpacs,superscriptaddress,amssymb]{revtex4}
\date{\today}

\usepackage{graphicx,threeparttable,amssymb}
\usepackage{dcolumn}% Align table columns on decimal point
\usepackage{bm}% bold math
\usepackage{color}
\begin{document}

\title{Shell-model study of boron, carbon, nitrogen and oxygen isotopes
based on monopole-based-universal interaction}

\author{Cenxi Yuan}
\email{cxyuan@pku.edu.cn} \affiliation{State Key Laboratory of
Nuclear Physics and Technology, School of Physics, Peking
University, Beijing 100871, China} \affiliation{Department of
Physics, University of Tokyo, Hongo, Bunkyo-ku, Tokyo 113-0033,
Japan}
\author{Toshio Suzuki}
\email{suzuki@phys.chs.nihon-u.ac.jp} \affiliation{Department of
Physics, College of Humanities and Sciences, Nihon University,
Sakurajosui 3, Setagaya-ku, Tokyo 156-8550, Japan}
\affiliation{National Astronomical Observatory of Japan, Mitaka,
Tokyo 181-8588, Japan }
\author{Takaharu Otsuka}
\email{otsuka@phys.s.u-tokyo.ac.jp} \affiliation{Department of
Physics, University of Tokyo, Hongo, Bunkyo-ku, Tokyo 113-0033,
Japan} \affiliation{Center for Nuclear Study, University of Tokyo,
Hongo, Bunkyo-ku, Tokyo 113-0033, Japan} \affiliation{National
Superconducting Cyclotron Laboratory, Michigan State University,
East Lansing, Michigan, 48824, USA}
\author{Furong Xu}
\email{frxu@pku.edu.cn} \affiliation{State Key Laboratory of Nuclear
Physics and Technology, School of Physics, Peking University,
Beijing 100871, China} \affiliation{Center for Theoretical Nuclear
Physics, National Laboratory for Heavy Ion Physics, Lanzhou 730000,
China}
\author{Naofumi Tsunoda}
\affiliation{Department of Physics, University of Tokyo, Hongo, Bunkyo-ku, Tokyo 113-0033, Japan}

\begin{abstract}
We study boron, carbon, nitrogen and oxygen isotopes with a newly
constructed shell-model Hamiltonian developed from
monopole-based-universal interaction ($V_{MU}$). The present
Hamiltonian can reproduce well the ground-state energies, energy
levels, electric quadrupole properties and spin properties of these
nuclei in full $psd$ model space including $(0-3)\hbar\omega$
excitations. Especially, it correctly describes the drip lines of
carbon and oxygen isotopes and the spins of the ground states of
$^{10}$B and $^{18}$N while some former interactions such as WBP and
WBT fail. We point out that the inclusion of $2\hbar\omega$
excitations is important in reproducing some of these properties. In
the present $(0+2)\hbar\omega$ calculations small but constant $E2$
effective charges appear to work quite well. As the inclusion of the
$2\hbar\omega$ model space makes rather minor change, this seems to
be related to the smallness of $^{4}$He core. Similarly, the spin g
factors are very close to free values. The applicability of tensor
and spin-orbit forces in free space, which are taken in the present
Hamiltonian, is examined in shell model calculations.
\end{abstract}


\maketitle

\section{\label{sec:level1}Introduction}
The existence of the unexpected doubly magic nucleus $^{24}$O shows
the exotic property of drip-line nuclei, that is, the change of
magic numbers far from the stability~\cite{janssens2009}. One of the
aims of theoretical works on nuclear structure is to describe both
stable nuclei and nuclei far from the stability in a unified
framework. In shell-model studies, for many of existing conventional
interactions, it is difficult to reproduce simultaneously the drip
lines of carbon and oxygen isotopes as well as some other properties
such as energies of $2_{1}^{+}$ states and $B(E2)$. From a
microscopic study, the inclusion of effects of three-body force is
important in describing the drip line of oxygen
isotopes~\cite{otsuka20102}. It is emergent to construct new
shell-model interactions applicable from the $\beta$-stability line
to the drip lines.

The realistic nucleon-nucleon (NN) interactions need to be
renormalized when applying to shell-model calculations because of
the short-range correlation and in-medium effect~\cite{dean2004}. NN
interaction is composed of three components, central force,
spin-orbit force and tensor force. Recent studies show that the
monopole components of tensor force barely change after the
renormalization and that the multipole components also change
little~\cite{otsuka2010,tsunoda2011}. Based on these studies, a
monopole-based-universal interaction ($V_{MU}$) including the bare
$\pi+\rho$ tensor force is introduced to describe the shell
evolution~\cite{otsuka2010}. As this $V_{MU}$ is constructed based
on monopole properties, it requires examination as to whether
$V_{MU}$ can be used in actual shell-model calculations or not. In
this paper we try to apply the $V_{MU}$ to shell-model calculations
in $psd$ model space.

In the $psd$ region, several effective interactions have been
introduced in shell-model calculations, such as PSDMK~\cite{mk1975},
WBT~\cite{wbt1992}, WBP~\cite{wbt1992} and SFO~\cite{suzuki2003}.
PSDMK, WBT and WBP interactions are all constructed in
$(0-1)\hbar\omega$ model space, which means that $0-1$ nucleons are
allowed to be excited from $p$ shell to $sd$ shell. The mixing
between $(0-1)\hbar\omega$ states and $(2-3)\hbar\omega$ states is
not considered in the fitting of the interaction. SFO, which
includes the $(2-3)\hbar\omega$ states, concentrates mostly on the
spin properties such as magnetic moments and Gamow-Teller
transitions. Up to now, the $\langle pp|V|sdsd\rangle$ matrix
elements, which represent the interaction between $(0-1)\hbar\omega$
states and $(2-3)\hbar\omega$ states, have not been well studied. In
Ref. \cite{suzuki2008}, the tensor part of the $\langle
psd|V|psd\rangle$ matrix elements is taken to be that of the
$\pi+\rho$ meson exchange potential and spin properties of C
isotopes have been studied. Recently, the study of microscopic
derivation of the effective interaction for the shell model in two
major shells is in progress~\cite{tsunoda2012}. It would be
interesting to apply them to shell-model calculations in future.

In this paper we try to construct the effective interaction in the
$psd$ space based on $V_{MU}$ to describe ground-state energies,
energy levels, electric quadrupole properties and spin properties.
The $\langle psd|V|psd\rangle$ and $\langle pp|V|sdsd\rangle$ matrix
elements are obtained based on $V_{MU}$ while phenomenological
effective interactions are used for the $p$-shell and $sd$-shell
parts to maintain the good description of the phenomenology by these
interactions. Microscopic interactions have been obtained based on
G-matrix method with medium modification \cite{jensen1995},
similarity renormalization group (SRG) method \cite{bogner2003} and
coupled-cluster method \cite{hagen2007}. While they produce
interesting results, fully microscopic calculations have not been
successful, as far as a good agreement to experiment is concerned.
We restrict here to a more phenomenological approach based on
$V_{MU}$ to study the spectroscopic properties of the nuclei to be
discussed.

In the next section, we introduce a new Hamiltonian. Coulomb
correction and center-of-mass correction are discussed in
Sec.~\ref{sec:level3} and Sec.~\ref{sec:level4}, respectively. In
Sec.~\ref{sec:level5}, we discuss the ground-state energies and
energy levels. We present the results of electric quadrupole
properties and spin properties in Sec.~\ref{sec:level6} and
Sec.~\ref{sec:level7}, respectively. A summary is given in
Sec.~\ref{sec:level8}.

\section{\label{sec:level2}Hamiltonian}

\begin{figure*}
\includegraphics[scale=0.9]{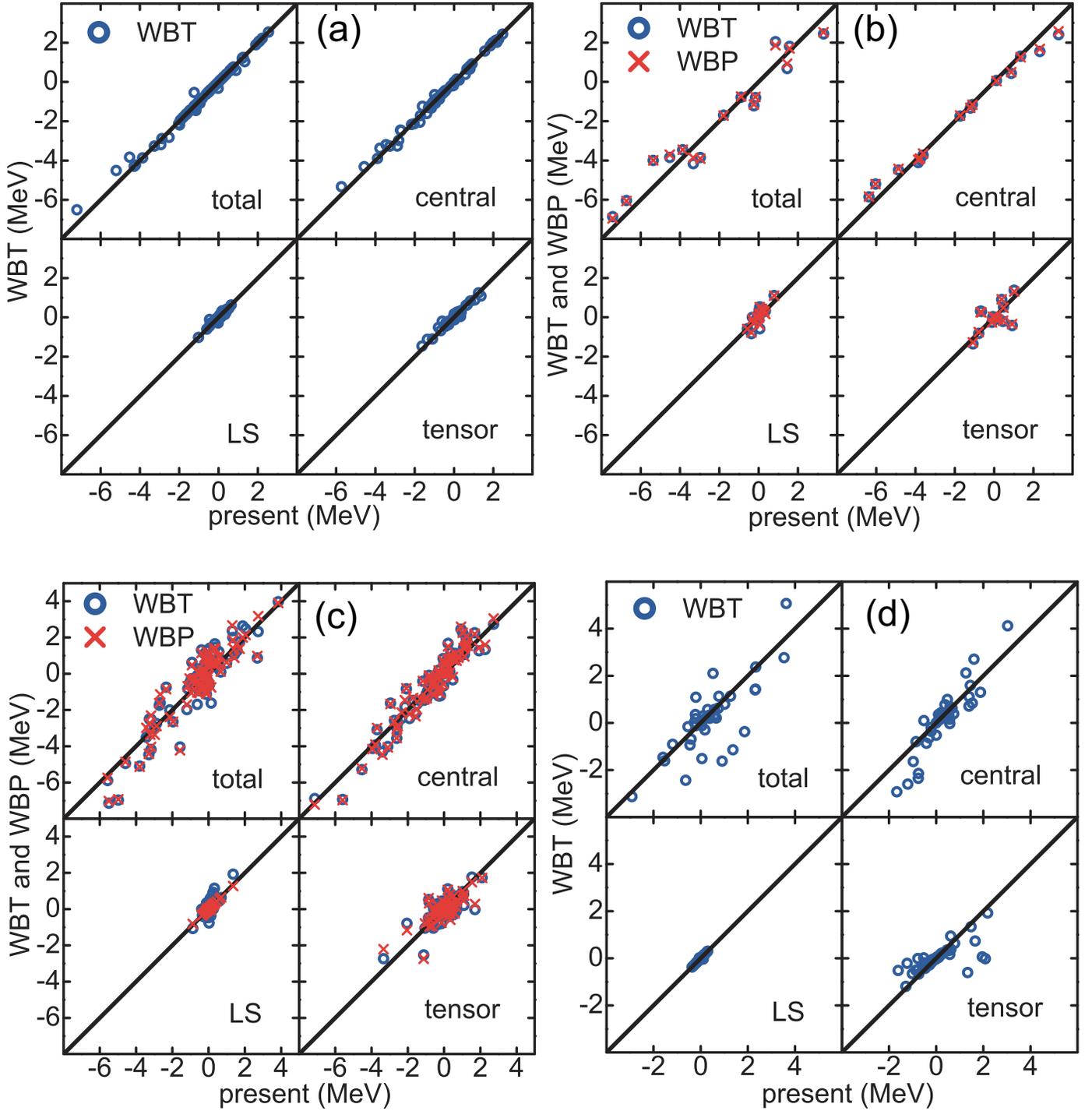}
\caption{\label{TBME}(Color online) Total TBME, TBME of central
force, spin-orbit force and tensor force in each part of
interaction: (a) $\langle sdsd|V|sdsd\rangle$, (b) $\langle
pp|V|pp\rangle$, (c) $\langle psd|V|psd\rangle$ and (d) $\langle
pp|V|sdsd\rangle$.}
\end{figure*}

\begin{figure*}
\includegraphics[scale=0.9]{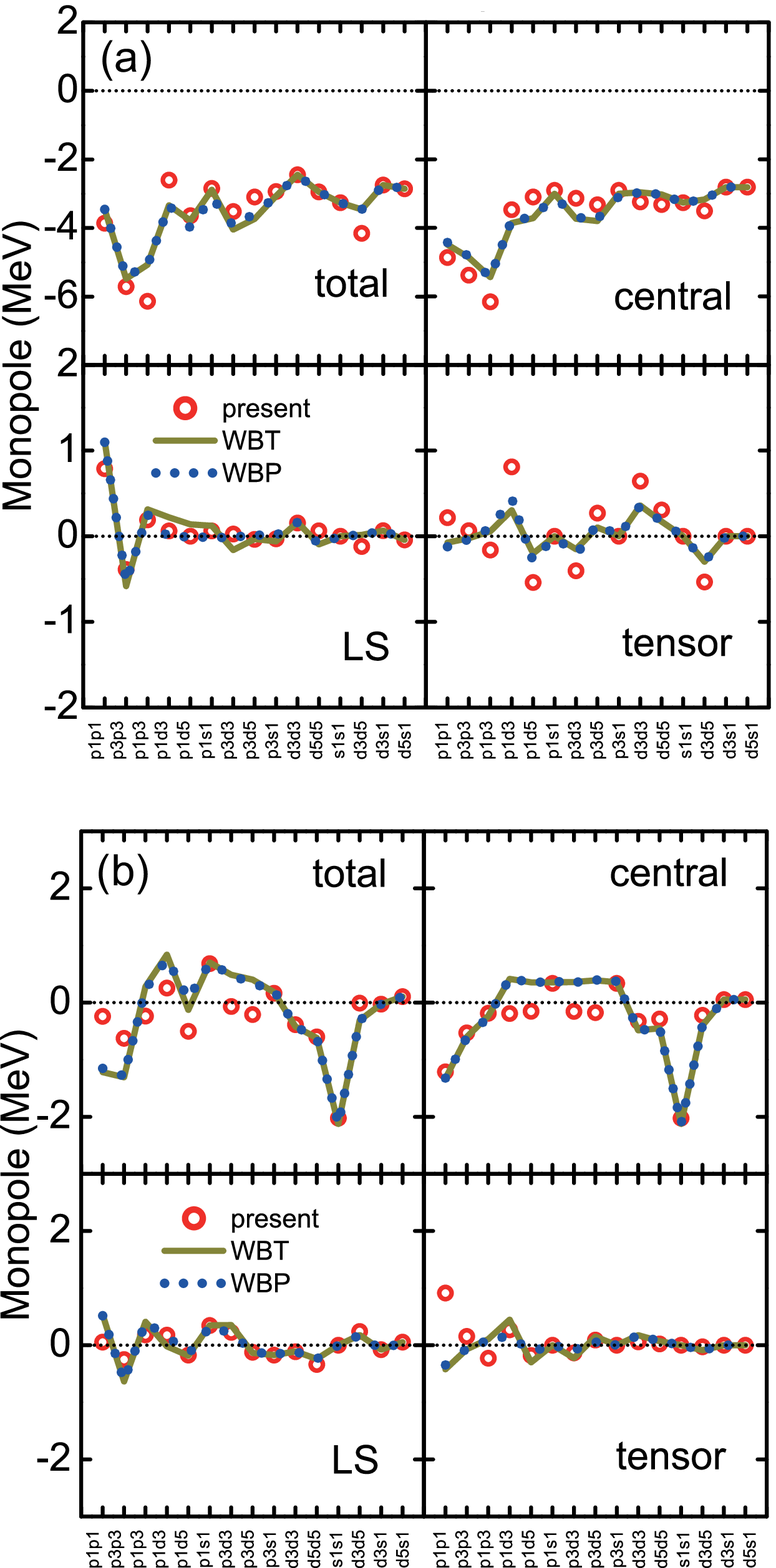}
\caption{\label{mono} (Color online) Monopole terms of total,
central, spin-orbit and tensor forces in each isospin channel: (a)
$T=0$ and (b) $T=1$.}
\end{figure*}

The present Hamiltonian is developed from $V_{MU}$, SFO and
SDPF-M~\cite{utsuno1999}. The two-body matrix elements (TBME) are
constructed as follows, $\langle pp|V|pp\rangle$ from SFO, $\langle
sdsd|V|sdsd\rangle$ from SDPF-M, $\langle psd|V|psd\rangle$ and
$\langle pp|V|sdsd\rangle$ from $V_{MU}$ plus spin-orbit force. In
the $\langle pp|V|pp\rangle$ matrix elements, we reduce the strength
of the monopole term $\langle
p_{1/2}p_{3/2}|V|p_{1/2}p_{3/2}\rangle_{T=0}$ by $0.5$ MeV from SFO.
This will improve the description of the ground-state energies of
these nuclei. The $\langle sdsd|V|sdsd\rangle$ matrix elements in
the present Hamiltonian are the same as SDPF-M. In earlier
interactions, such as WBP and WBT, the matrix elements $\langle
pp|V|sdsd\rangle$ are not considered in the fitting procedure. The
strength of the interaction in $\langle pp|V|sdsd\rangle$ in WBP and
WBT is the same as in $\langle psd|V|psd\rangle$ in WBP. In the
present interaction, strengths of these two parts of the interaction
are not taken to be the same. The $V_{MU}$ includes Gaussian type
central force and $\pi+\rho$ tensor force. We use M3Y~\cite{m3y1977}
force for the spin-orbit force. We keep the spin-orbit and tensor
forces unchanged. The form of the interactions in the matrix
elements of $\langle psd|V|psd\rangle$ and $\langle
pp|V|sdsd\rangle$ is as follows,
\begin{equation}\label{v}
 V=V_{central}+V_{spin-orbit}(\text{M3Y})+V_{tensor}(\pi+\rho),
\end{equation}
with $V_{central}$ being,
\begin{equation}\label{vc}
 V_{central}=\sum_{S,T} f_{S,T}P_{S,T}exp(-(r/\mu)^{2}),
\end{equation}
where $S~(T)$ means spin (isospin), $P_{S,T}$ is the projection
operator on $S, T$ channel. $r$ and $\mu$ are distance between two
nucleons and Gaussian parameter, respectively. $f_{ST}$ is the
strength of the central force. In the original $V_{MU}$,
$f_{0,0}=f_{1,0}=-166$ MeV, $f_{0,1}=0.6f_{0,0}$ and
$f_{1,1}=0.8f_{0,0}$~\cite{otsuka2010}. In the present study, we
reduce the central force in $\langle psd|V|psd\rangle$ and $\langle
pp|V|sdsd\rangle$ matrix elements by factors $0.85$ and $0.55$ from
the original $V_{MU}$, respectively. The final interaction in the
$\langle psd|V|psd\rangle$ ($\langle pp|V|sdsd\rangle$) matrix
elements is
\begin{eqnarray}
\label{vfinal}
V = 0.85(0.55)V_{central}+ \nonumber \\
 V_{spin-orbit}(\text{M3Y})+ V_{tensor}(\pi+\rho).
\end{eqnarray}
Notice that the spin-orbit force and the tensor force are kept
unchanged. The TBME are calculated with harmonic oscillator
parameter $\hbar\omega=45A^{-1/3}-25A^{-2/3}$ where $A=18$ which is
the average mass number of the investigated nuclei from $^{10}$B to
$^{26}$O. The $sd$-shell single-particle energies (SPE's) in SDPF-M
are $\epsilon_{d_{5/2}}=-3.95$ MeV, $\epsilon_{d_{3/2}}=1.65$ MeV
and $\epsilon_{s_{1/2}}=-3.16$ MeV, which takes $^{16}$O as the core
\cite{utsuno1999}. In the present shell-model calculations, $^{4}$He
is chosen as the core, thus the $sd$-shell SPE's in the present
Hamiltonian should be adjusted to give the same one-particle
excitation energies of $^{17}$O as in SDPF-M. The adjusted SPE's are
$\epsilon_{d_{5/2}}=8.01$ MeV, $\epsilon_{d_{3/2}}=10.11$ MeV and
$\epsilon_{s_{1/2}}=2.11$ MeV. The $p$-shell SPE's are obtained
based on SFO but with slight changes by fitting the ground-state
energies of the studied nuclei and related levels such as the
$1/2^{-}_{1}$ state in $^{11}$B and $3/2^{-}_{1}$ state in $^{13}$C.
We obtain $\epsilon_{p_{3/2}}=1.05$ MeV and
$\epsilon_{p_{1/2}}=5.30$ MeV. The detailed TBME of the present
Hamiltonian can be obtained by contacting the authors.

We compare the TBME of the present Hamiltonian with those of WBT and
WBP in Fig.~\ref{TBME}. The TBME of central, spin-orbit and tensor
interactions are also presented by the spin-tensor decomposition
method \cite{kirson1973}. The $sd$ and $ppsdsd$ parts of WBT and WBP
are the same between the two. So we show only the WBT result in
these two parts. The $sd$ part of the present interaction is from
SDPF-M which is modified from USD (the same as the $sd$ part of WBT)
interaction. There is not much difference between WBT and the
present interaction in the $sd$ part. In the $p$ part, all these
three interactions, present, WBT and WBP, are fitted to low-lying
levels of the $p$-shell nuclei. The difference among these three
interactions is not large except for the tensor force. In the $psd$
and $ppsdsd$ parts of the interaction, the deviation of the present
interaction from WBT (WBP) turns out to be larger. The central force
of the present interaction in the $psd$ part is $0.85V_{MU}$. We
find that this strength is proper as the number of points above the
diagonal line is close to that below the line as shown in the
Fig.~\ref{TBME}(c). It is interesting that the spin-orbit
interaction of the present interaction is very similar to that of
WBP interaction in both the $psd$ and $ppsdsd$ parts. In these two
parts of the interaction, WBP has $10$ parameters for the potential
fitting while the present interaction is taken from the M3Y
potential. Quite similar results between WBP and the present
interaction indicate that the spin-orbit force is rather well
determined compared to the central force.

Figure~\ref{mono} presents the monopole terms of the interactions
and their spin-tensor components. Monopole term is a weighted
average of TBME for orbits $j$ and $j'$~\cite{bf1964,zuker1981},
\begin{equation}\label{mono}
 V^{T}_{j,j'}=\frac{\sum_{J}(2J+1)\langle jj'|V|jj'\rangle_{J,T}}{\sum_{J}(2J+1)}.
\end{equation}
The monopole terms are presented in three groups, $pp$, $sdsd$ and
$psd$ in each picture. In each group the central monopole is
relatively flat compared with the total monopole. The total
interaction can be recognized as a global central force plus other
staggers. The $T=0$ central monopole is the most attractive among
all these six central, spin-orbit and tensor monopoles. The nuclear
binding energy comes mostly from this interaction. Both $T=0$ and
$T=1$ spin-orbit monopoles of the present interaction are very close
to those of WBP. This is consistent with the analysis of the
spin-orbit part of the TBME. Comparing with WBT, the present
spin-orbit monopoles are also not much different. The present tensor
force is stronger than WBT and WBP in $T=0$ channel, more attractive
in $\langle p_{1/2}d_{5/2}|V|p_{1/2}d_{5/2}\rangle$ and more
repulsive in $\langle p_{3/2}d_{5/2}|V|p_{3/2}d_{5/2}\rangle$. In
the $sd$ region of nuclei, this effect of the tensor force is
canceled as the $p_{1/2}$ and $p_{3/2}$ orbits are fully
occupied~\cite{otsuka2005}. Going to $psd$ region, such as neutron
rich boron, carbon and nitrogen isotopes, the opposite sign of the
monopoles of the tensor force turns out to be important.

\section{\label{sec:level3}Coulomb correction}
In the shell-model study, Coulomb interaction is not included in
many cases, in order to keep the isospin symmetry. When we compare
the ground-state energies between theoretical results and observed
values, Coulomb correction is needed. Present calculations in $psd$
model space do not include the ground-state energies of $^{4}$He,
$E(^{4}He)$, which also needs to be removed. The total correction is
as follow,
\begin{equation}\label{coulomb}
E_{correction}=E_{exp.}-E_{Coulomb}-E(^{4}He),
\end{equation}
where $E_{Coulomb}$ and $E_{correction}$ are the energy of Coulomb
correction and the ground-state energy after the correction,
respectively. $E(^{4}He)=-28.296$ MeV. $E_{Coulomb}$ is calculated
through similar method used in the construction of WBT and WBP
interactions~\cite{wbt1992}. We calculate the energy difference of
mirror nuclei near $N=Z$ where the observed ground-state energies
are taken from~\cite{audi2003}. This $E_{Coulomb}$ is dependent only
on Z in our calculation. $E_{Coulomb} = 1.075
~(Z=3),~2.720~(Z=4),~4.593~(Z=5),~7.368~(Z=6),~10.248~(Z=7),~13.854
~(Z=8)$~MeV.

\begin{figure}
\includegraphics[scale=0.3]{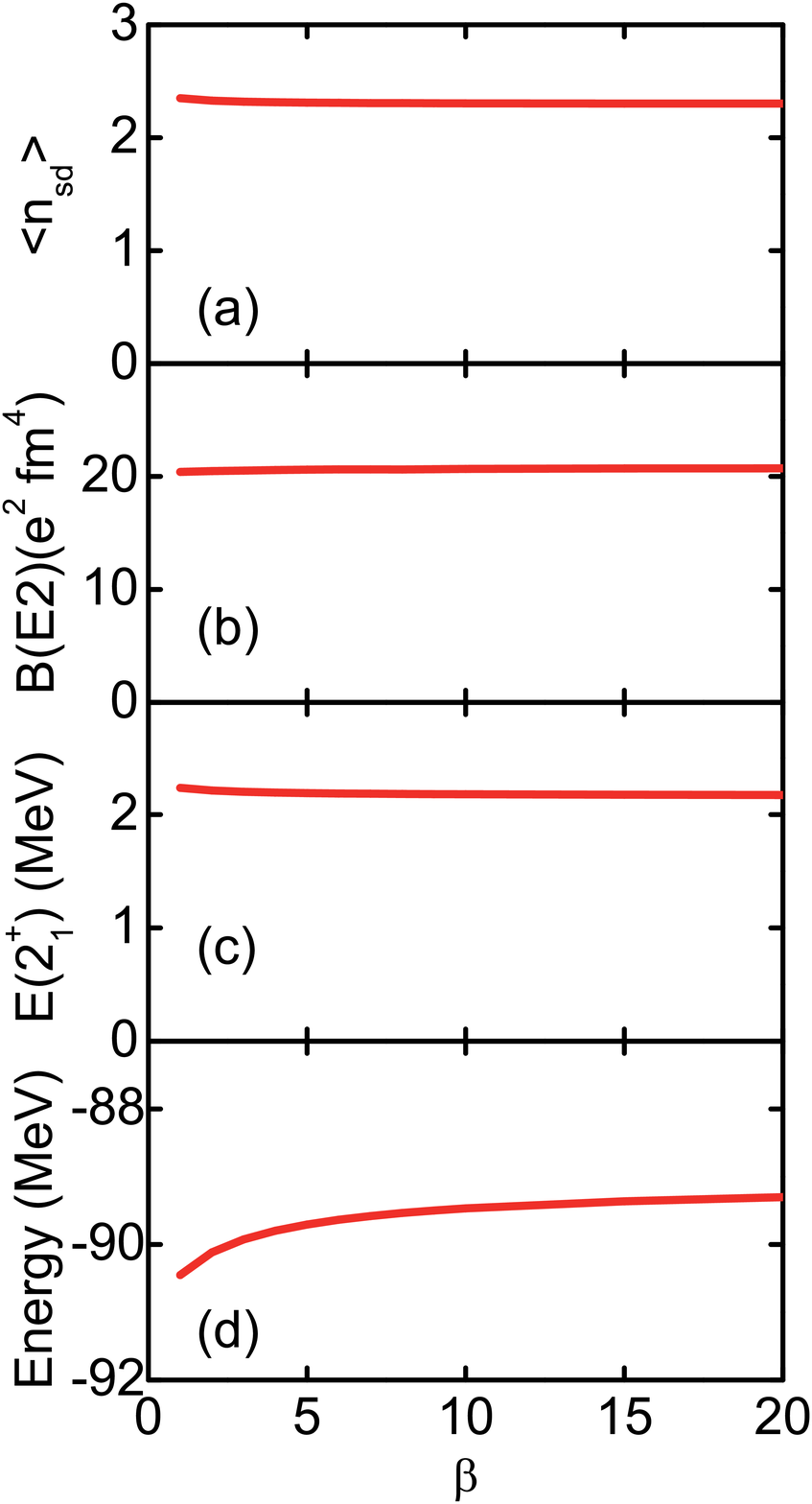}
\caption{\label{cm} (Color online) Effects of the center-of-mass
corrections in four physical quantities: (a) nucleon number in
$sd$-shell, (b) $B(E2;0_{1}^{+}\rightarrow 2_{1}^{+})$ value, (c)
$E(2_{1}^{+})$ and (d) ground-state energy of $^{16}$C. Their values
change as the function of $\beta$.}
\end{figure}

\section{\label{sec:level4}Center-of-mass correction}

As our calculation is done in two major shells, we need
center-of-mass (c.m.) correction to remove the spurious components
which come from the c.m. motion. We use the method suggested by
Gloeckner and Lawson~\cite{cm1974}. In the calculations, Hamiltonian
is $H'=H_{SM}+\beta H_{c.m.}$, where $H_{SM}$ and $H_{c.m.}$ are
original and c.m. Hamiltonians, respectively. If $\beta$ is large
enough, the effect of the c.m. motion is small enough in low lying
states. Figure~\ref{cm} indicates some physical quantities of
$^{16}$C to check whether this method works or not, and how large
$\beta$ is needed.

We find that the number of nucleons in the $sd$-shell,
$B(E2;0_{1}^{+}\rightarrow2_{1}^{+})$ or the energy of $2_{1}^{+}$
in $^{16}$C hardly change when $\beta$ changes. The ground-state
energy of $^{16}$C changes quickly when $\beta$ is small. For
$\beta>10$, it becomes almost flat. We use $\beta=10$ in the
following calculations.

\section{\label{sec:level5}Ground-state energy and energy level}

\begin{figure}
\includegraphics[scale=0.35]{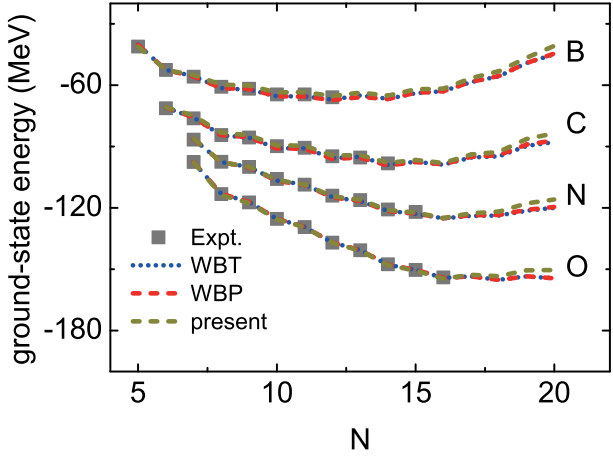}
\caption{\label{BE} (Color online) Ground-state energies of boron,
carbon, nitrogen and oxygen isotopes. Experimental values are taken
from Ref.~\cite{audi2003}}
\end{figure}

\begin{figure}
\includegraphics[scale=0.35]{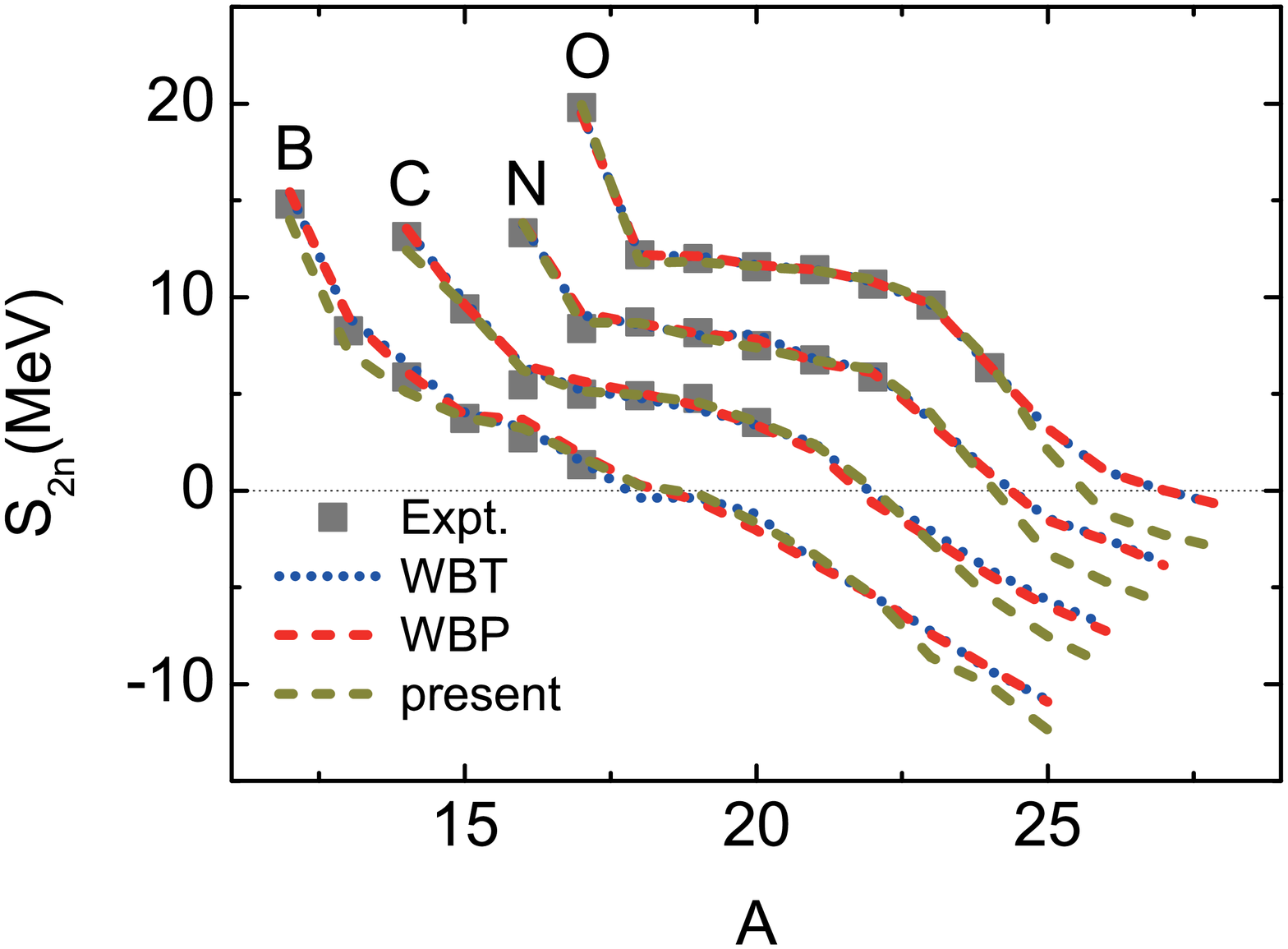}
\caption{\label{S2n} (Color online) Two neutron separation energies,
$S_{2n}$, of boron, carbon, nitrogen and oxygen isotopes.
Experimental values are taken from Ref.~\cite{audi2003}}
\end{figure}

\begin{figure}
\includegraphics[scale=0.47]{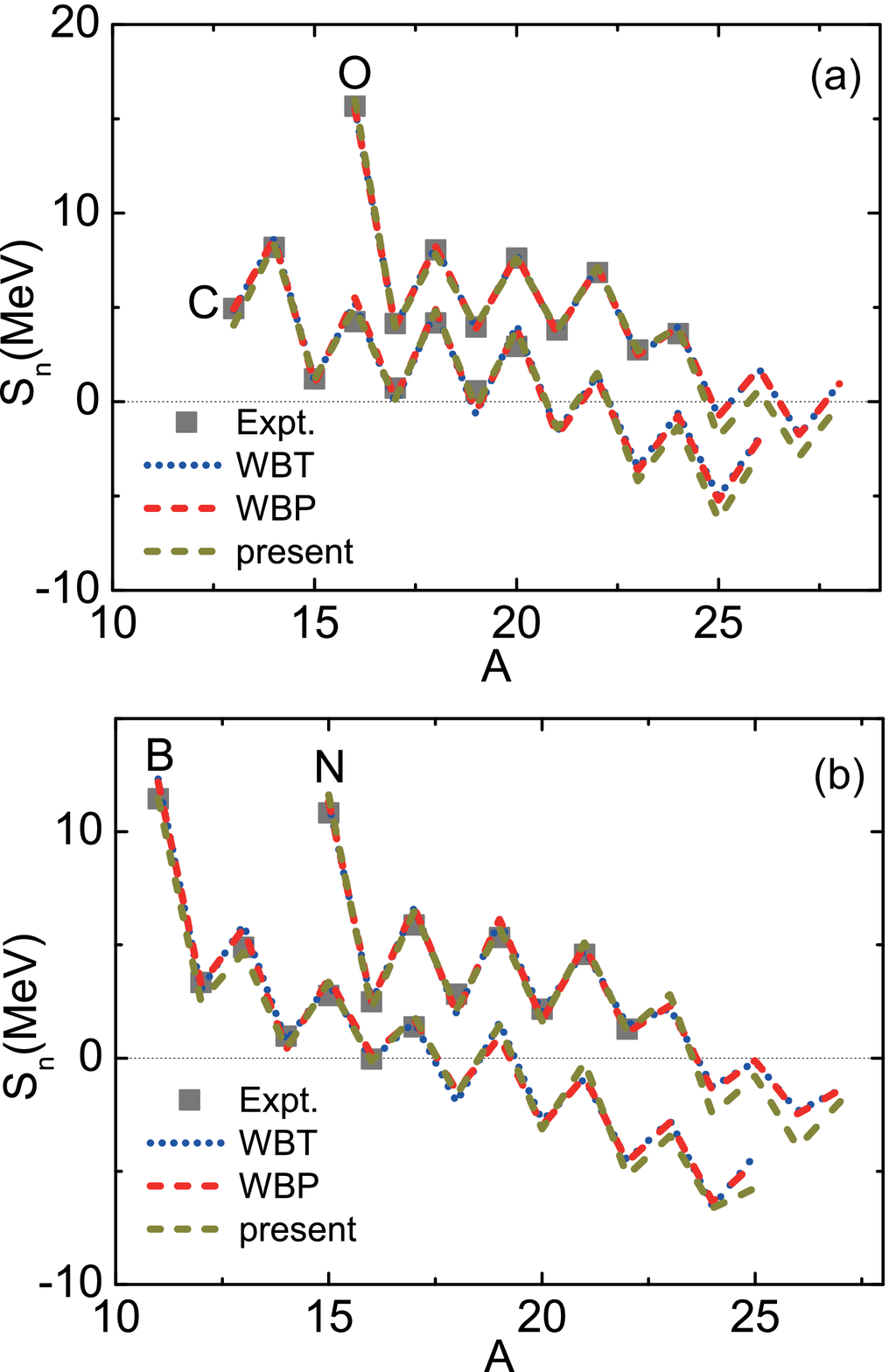}
\caption{\label{Sn} (Color online) One neutron separation energies,
$S_{n}$ of boron, carbon, nitrogen and oxygen isotopes. Experimental
values are taken from Ref.~\cite{audi2003}.}
\end{figure}

\begin{figure}
\includegraphics[scale=0.35]{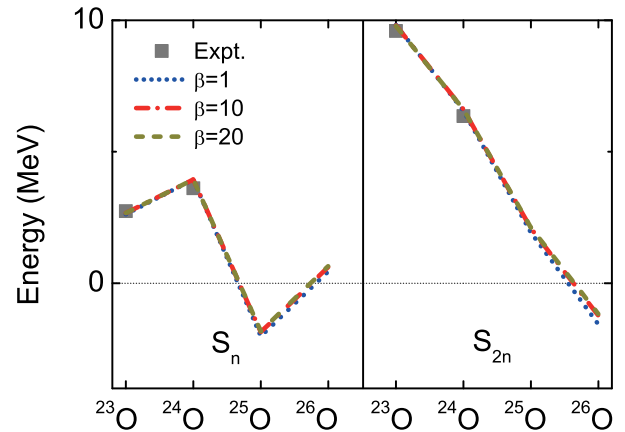}
\caption{\label{beta} (Color online) Comparison of $S_{n}$ and
$S_{2n}$ under three different values of the center-of-mass
parameter, $\beta=1,~10$ and $20$.}
\end{figure}

\begin{figure*}
\includegraphics[scale=0.9]{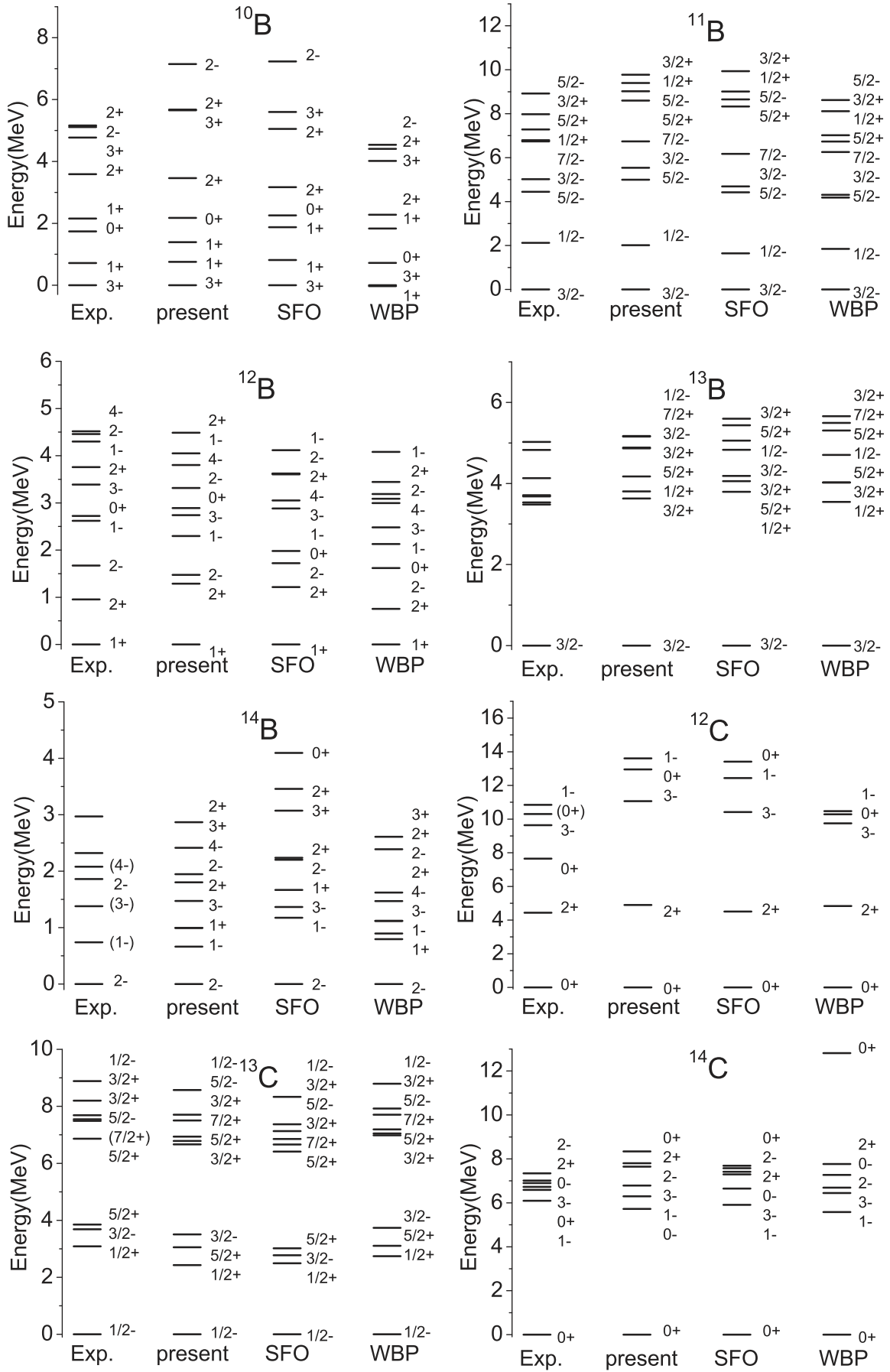}
\caption{\label{B} Energy levels for boron and light carbon
isotopes, obtained in the present, SFO and WBP calculations,
compared with experimental
data~\cite{nndc,stanoiu2008,sohler2008,strongman2009,stanoiu2004,schiller2007}.}
\end{figure*}

\begin{figure*}
\includegraphics[scale=0.9]{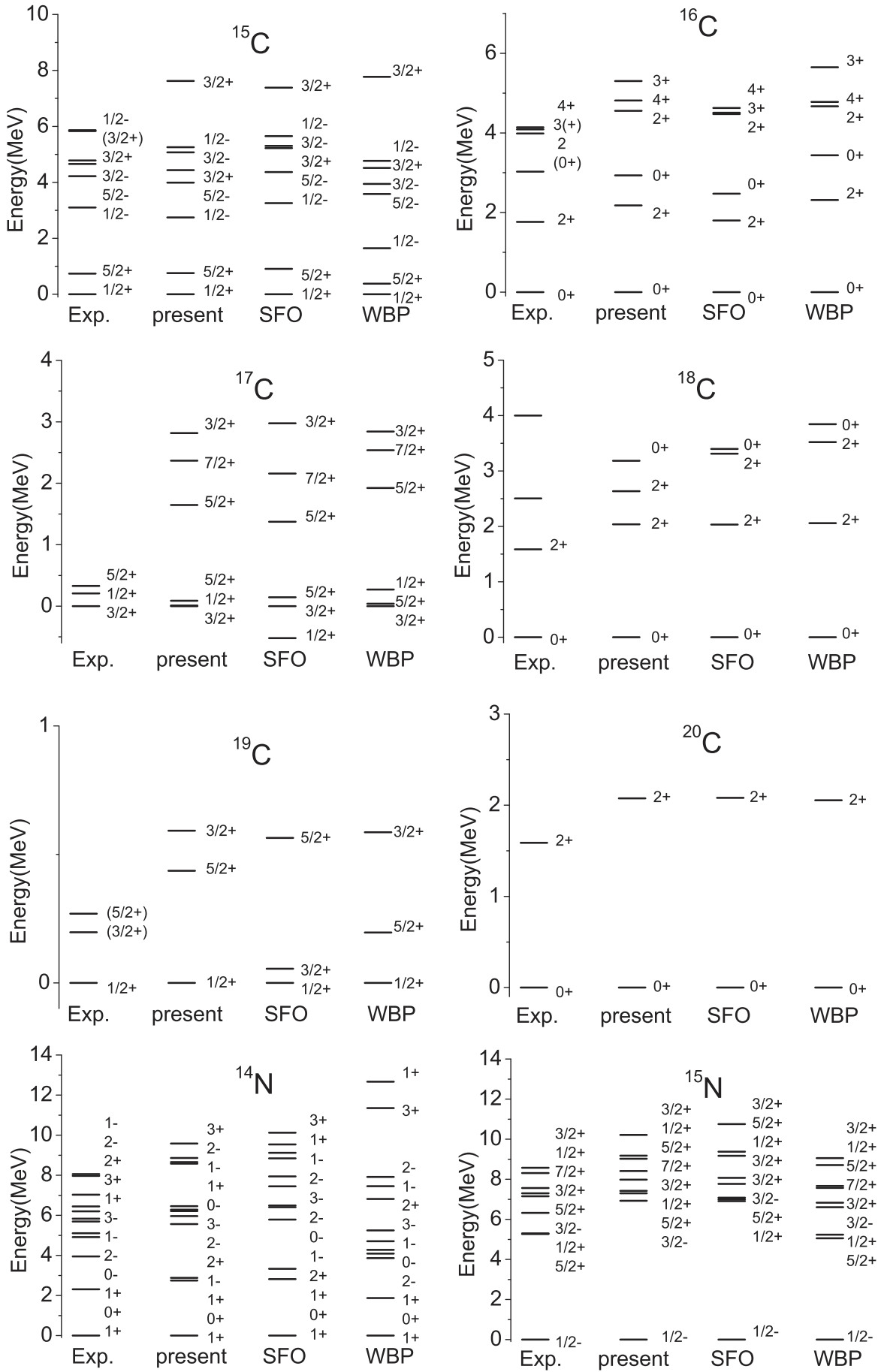}
\caption{\label{C} Similar to Fig.~\ref{B}, but for heavier carbon
and light nitrogen isotopes.}
\end{figure*}

\begin{figure*}
\includegraphics[scale=0.9]{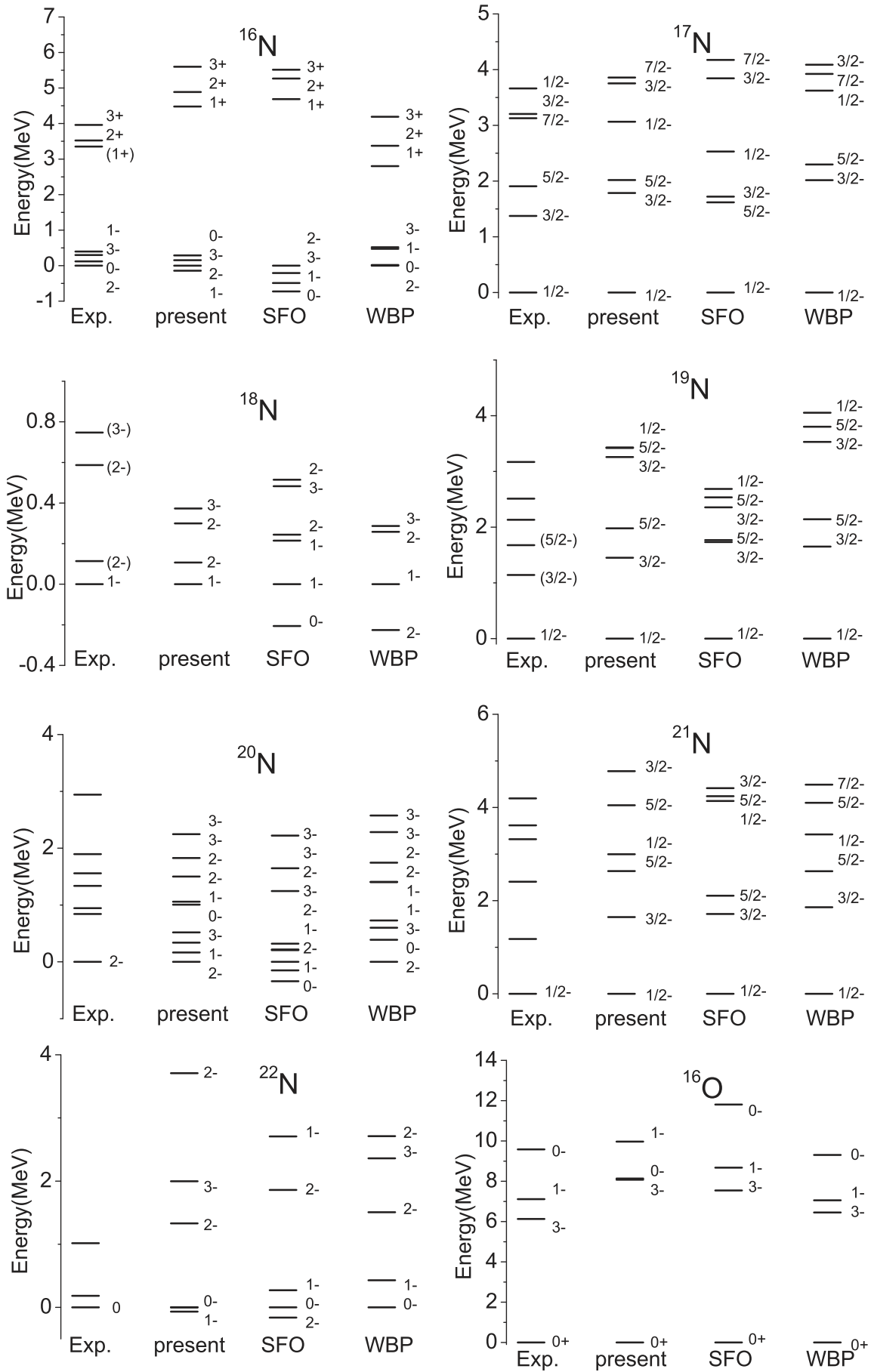}
\caption{\label{N} Similar to Fig.~\ref{B}, but for heavier nitrogen
and light oxygen isotopes.}
\end{figure*}

\begin{figure*}
\includegraphics[scale=0.9]{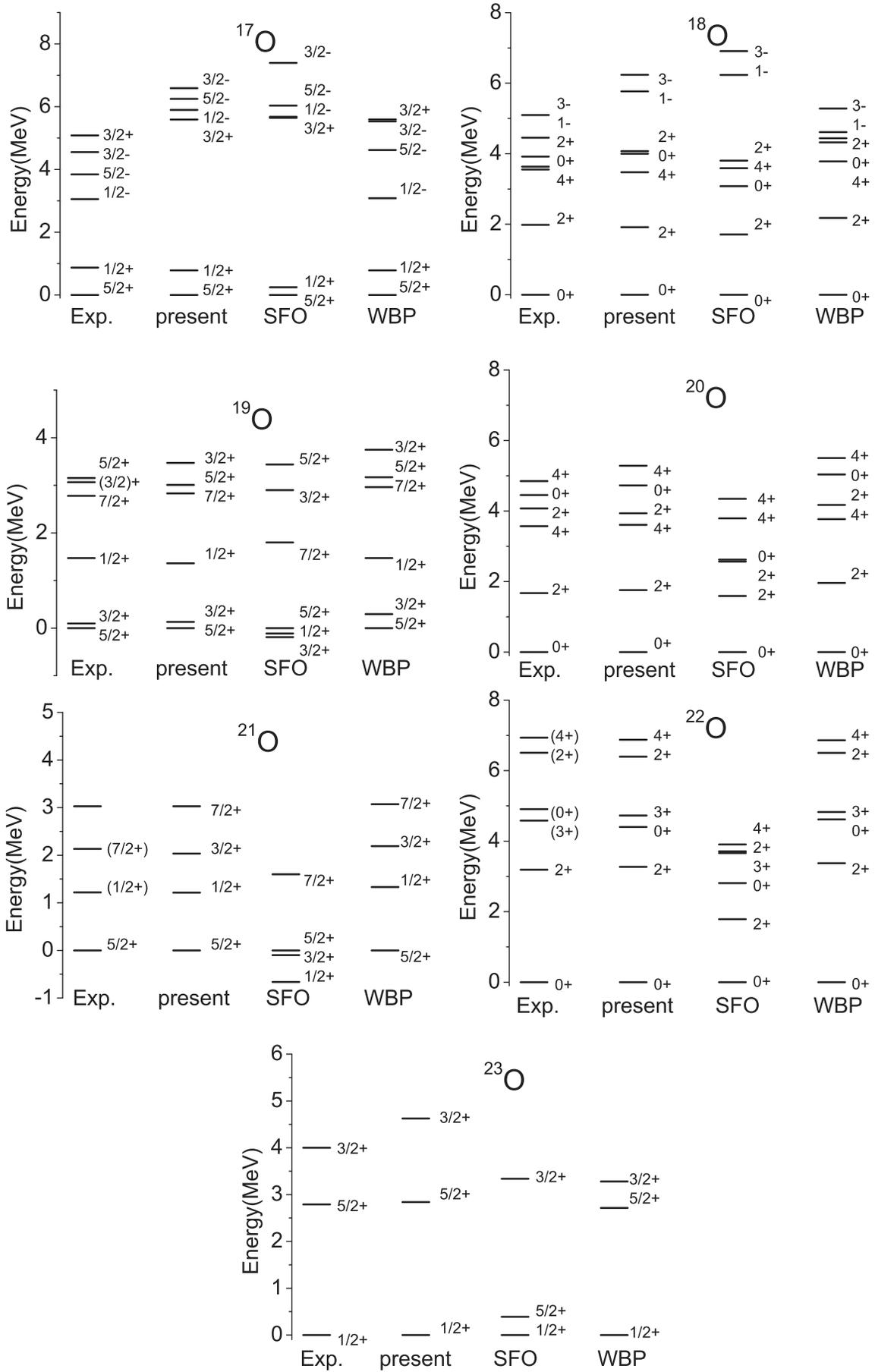}
\caption{\label{O} Similar to Fig.~\ref{B}, but for heavier oxygen
isotopes.}
\end{figure*}

\begin{figure}
\includegraphics[scale=0.32]{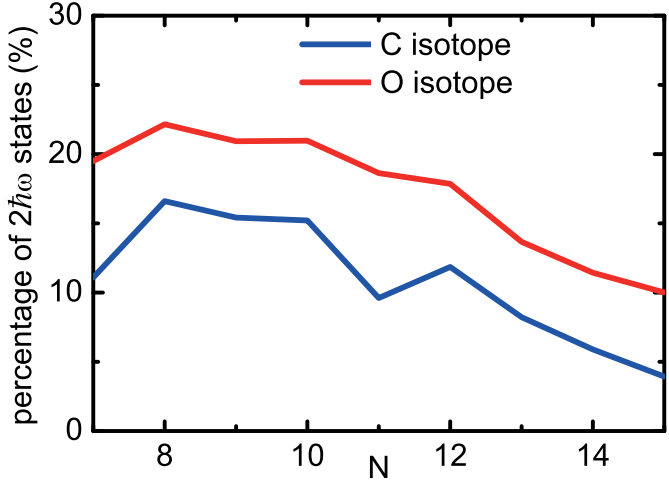}
\caption{\label{2hw} (Color online) Percentage of $2\hbar\omega$
components in carbon and oxygen isotopes.}
\end{figure}

\begin{figure}
\includegraphics[scale=0.57]{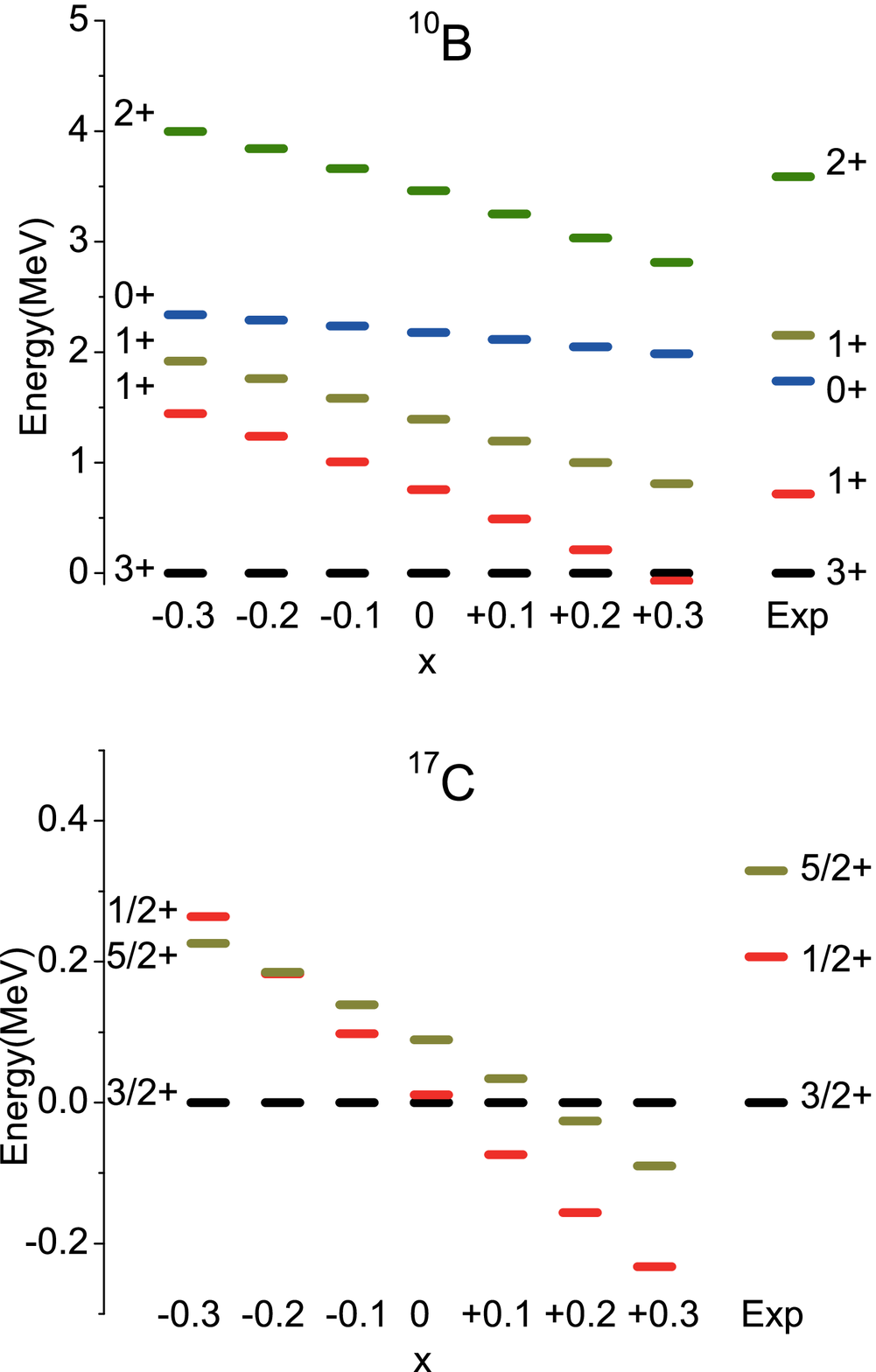}
\caption{\label{10B17C2hw} (Color online) Energy levels of $^{10}$B
and $^{17}$C as the function of x which specifies the strength of
the central force, $\langle
pp|V|sdsd\rangle(central)=(0.55+x)V_{MU}(central)$}
\end{figure}

The nuclei $^{22}$C, $^{23}$N and $^{24}$O are the last bound nuclei
in neutron rich side of C, N and O isotopes~\cite{thoennessen2012}.
The neutron-drip lines in elements beyond oxygen is not determined
yet~\cite{thoennessen2012}. In WBT and WBP, $^{22}$C is unbound and
$^{26}$O is bound. The present Hamiltonian improves the description
of drip lines of C and O isotopes. Figures~\ref{BE} to~\ref{Sn}
present the ground-state energies as well as one and two neutron
separation energies $S_{n}$ and $S_{2n}$, for B, C, N and O
isotopes. In O isotopes, the WBT and WBP have the same result as
their $sd$ parts. From $S_{n}$ and $S_{2n}$ of O isotope, one can
see that both WBT and the present Hamiltonian predict $^{25}$O to be
unbound consistently with experiment~\cite{thoennessen2012}. The
$^{26}$O is about $1.2$ MeV unbound in the present result and $1.0$
MeV bound in WBT. The positive $S_{n}$ value in $^{26}$O indicates
that $^{26}$O is one-neutron bound but two-neutron unbound. In N
isotopes, all these three interactions can reproduce that $^{23}$N
is bound and $^{24}$N and $^{25}$N are unbound. $^{21}$C is unbound
experimentally~\cite{thoennessen2012} and also unbound in the
calculations by all these three interactions. $^{22}$C is $0.1$ MeV
bound in the present result and $0.2$ MeV and $0.6$ MeV unbound in
WBT and WBP, respectively. $^{16}$B is $40(60)$ keV
unbound~\cite{audi2003}. It is $144$ keV bound in WBP and $65$ keV
and $153$ keV unbound in WBT and the present Hamiltonian,
respectively. All Hamiltonians succeed in describing unbound
$^{18}$B. In the experiment~\cite{thoennessen2012}, WBP and present
Hamiltonian, it is one-neutron unbound. But in WBT it is both one
and two-neutron unbound. $^{19}$B, which is experimentally
bound~\cite{thoennessen2012}, is unbound with $160$ keV, $381$ keV
and $538$ keV in the present Hamiltonian, WBT and WBP, respectively.

Here we briefly summarize the descriptions of drip lines by these
three Hamiltonians. The present Hamiltonian is successful in
describing all drip-line nuclei except for $^{19}$B while WBT fails
in $^{26}$O, $^{22}$C, $^{18}$B and $^{19}$B and WBP fails in
$^{26}$O, $^{22}$C, $^{16}$B and $^{19}$B. One reason that the
present interaction improves the description of drip lines is the
inclusion of the mixing between $0\hbar\omega$ and $2\hbar\omega$
configurations. WBT and WBP have mass-dependent term in the
$sd$-shell~\cite{wbt1992}. Going from $^{18}$O to $^{28}$O, the $sd$
shell interaction decreases, which makes the nuclei less binding. We
find that the mixing between $0\hbar\omega$ and $2\hbar\omega$
states has a similar effect. Partial effect of mass dependence
therefore comes from the mixing between $0\hbar\omega$ and
$2\hbar\omega$ states which is not included in WBT and WBP. We will
discuss more about the contribution of $2\hbar\omega$ states later.

In order to see if the prediction on the neutron drip line is
sensitive to the center-of-mass parameter, $\beta$, we have made
calculations with assuming three different values of the parameter.
Figure~\ref{beta} displays the calculations of one- and two-neutron
separation energies under different $\beta$ values, showing that the
value of $\beta=10$ used in the present work is large enough to
remove the spurious center-of-mass components. For example, the
neutron separation energies of neutron-rich oxygen isotopes change
about $100$~keV when increasing $\beta$ from $1$ to $10$, while the
separation-energy variation is about $20$~keV when increasing
$\beta$ from $10$ to $20$. It is consistent to discussions in
Sec.~\ref{sec:level4} that the physical properties are well
convergent when $\beta=10$.

Figures~\ref{B} to \ref{O} present the energy levels of B, C, N, and
O isotopes. The agreement between experiment and the present work is
fairly good. Especially for $^{10}$B and $^{18}$N, we can reproduce
the spins of the ground states of these two nuclei while WBP and WBT
fail. WBT also fails in describing the spins of the ground states of
$^{17}$C, $^{19}$C and $^{16}$N. We only show WBP results here as
WBP is similar to WBT and is better in the description of the spins
of the ground states. The ground states of $^{16}$N and $^{22}$N are
about $100$ keV higher in the present interaction. The SFO can also
reproduce the spin of the ground state of $^{10}$B and nuclei
nearby. But it fails in some neutron rich nuclei such as $^{19}$O
and $^{21}$O. This is because the $\langle sdsd|V|sdsd\rangle$ part
of SFO is from renormalized G-matrix~\cite{suzuki2003}. The
interaction is too attractive without the contribution of three-body
forces~\cite{otsuka20102}. The first $1/2^{+}$ and $5/2^{+}$ states
in $^{19}$O, $^{21}$O and $^{23}$O indicate that the neutron
$1s_{1/2}$ orbit is too low compared with neutron $0d_{5/2}$ orbit
in SFO. This situation can be improved by using effective
interactions such as SDPF-M or including the contribution of
three-body forces~\cite{otsuka20102}, for instance.

The energy difference between first $3^{+}$ and $1^{+}$ in $^{10}$B
can be reproduced well by both SFO and the present Hamiltonian. This
is partly because the $0p_{1/2}$ orbit is much higher than
$0p_{3/2}$ orbit and partly because the strength of $\langle
pp|V|sdsd\rangle$ is chosen properly which will be discussed later.
For $^{10}$B and $^{11}$B, an \emph{ab initio} no-core shell model
calculation based on chiral perturbation theory showed that the
inclusion of the three-body force is necessary to reproduce the
ground-state spins ~\cite{navratil2007}. In the present Hamiltonian,
the phenomenological effective two-body interaction is mostly
obtained by fitting experimental data. Therefore, the effective
interaction obtained thus includes, at least partly, the three-body
effect. Our calculations show this equivalence. In
Ref.~\cite{otsuka20102}, it was pointed out that \emph{ab initio}
interaction without three-body force cannot reproduce the neutron
drip line of oxygen isotopes, while \emph{ab initio} interaction
with three-body force or phenomenological two-body interactions may
describe the drip line.

The $\nu(sd)^{3}$ configuration shows different structure in $N=11$
isotones from $^{17}$C to $^{19}$O~\cite{wiedeking2008}.
$\nu(0d_{5/2})^{3}$ can couple to $J=5/2$ with seniority $v=1$ or
couple to $J=3/2$ with seniority $v=3$. The structure of
$\nu(sd)^{3}$ as well as the low lying states in $N=11$ isotones is
a subtle problem because of these two configurations together with
$\nu(0d_{5/2})^{2}(1s_{1/2})^{1}$, $\nu(0d_{5/2})^{1}(1s_{1/2})^{2}$
and many small components. For the first time, the present
Hamiltonian in full $psd$ model space reproduces the low lying
states in all these three nuclei, $^{17}$C, $^{18}$N and  $^{19}$O.
$\langle (0d_{5/2})^{2}|V|(0d_{5/2})^{2}\rangle_{J=0~T=1}$ paring
interaction contributes to this good agreement because this paring
is reduced in SDPF-M which will make the $v=1$ state less bound and
keep $v=3$ state unchanged. As one can see, $2^{+}$ state in
$^{18}$N, $5/2^{+}$ states in $^{19}$O and $^{17}$C in the present
results become higher compared with those in WBP results. Other
matrix elements also contribute to this subtle problem. We will
discuss the contribution of $\langle pp|V|sdsd\rangle$ in $^{17}$C
later.

The WBP and WBT results show more expanded energy levels compared
with observed energy levels in C and N
isotopes~\cite{stanoiu2008,sohler2008}. This can be improved by
reducing neutron-neutron interactions by $25\%$ (for C isotopes) or
$12.5\%$ (for N isotopes) in the $sd$-shell in WBP and
WBT~\cite{stanoiu2008,sohler2008}. The spectra of the present
interaction are not so expanded as in WBP and WBT for C, N and O
isotopes.

In the present work, energy levels of unnatural parity state are not
fully considered. One reason is that experimental data of these
energy levels are not much available in neutron rich nuclei. Another
reason is that the dimension of the calculation increases quickly
when including $3\hbar\omega$ components. We can improve the
description of these unnatural parity states with more experimental
data and more advanced computers in the future.

The strength of the interaction in the $\langle pp|V|sdsd\rangle$
matrix elements is not determined in PSDMK, WBP and WBT and not
fully considered in SFO. In the present interaction, the strength of
the central part of $\langle pp|V|sdsd\rangle$ is $55\%$ of
$V_{MU}$. We will show some examples that the $\langle
pp|V|sdsd\rangle$ matrix elements are important in describing the
nuclei being studied. The total wave function of a nucleus can be
written as $\Psi=a\Psi(0\hbar\omega)+b\Psi(2\hbar\omega)$.
Figure~\ref{2hw} shows the probability, $b^{2}$, of $2\hbar\omega$
component. It is clear that the probability $b^{2}$ is very
sensitive to neutron numbers. When the neutron number increases from
$8$ to $15$, the value of $b^{2}$ decreases except a singular point
$^{17}$C. In WBP and WBT, the $sd$ part includes the mass-dependent
term $(18/A)^{0.3}$~\cite{wbt1992}. Only with this effect, WBP and
WBT can reproduce well the ground-state energies of these nuclei.
The mass-dependent term is needed for calculations of nuclei in a
large mass range because the nuclear force is related to the radii
of nuclei as well as the nucleon number $A$. But in a range of
nuclei with small mass numbers, the effect of mass dependence is not
obvious when we include $2\hbar\omega$ components. In the present
Hamiltonian, we can reproduce well ground-state energies, separation
energies and energy levels of B, C, N and O isotopes without mass
dependent term. One can see from Fig.~\ref{2hw} that the inclusion
of $2\hbar\omega$ components will automatically contain a part of
mass-dependent effects. More works are needed to study the mass
dependent effects in light nuclei.

The $\langle pp|V|sdsd\rangle$ matrix elements are also important
for energy levels in certain nuclei. Figure~\ref{10B17C2hw} shows
the dependence of the energy levels in $^{10}$B and $^{17}$C on the
interaction. Energy differences, such as difference between
$3^{+}_{1}$ and $1^{+}_{1}$ in $^{10}$B and that between
$3/2^{+}_{1}$ and $5/2^{+}_{1}$ in $^{17}$C, are very sensitive to
the strength of $\langle pp|V|sdsd\rangle$. The energy difference
between $3^{+}_{1}$ and $0^{+}_{1}$ in $^{10}$B, on the other hand,
is hardly changed when the strength of the central part of $\langle
pp|V|sdsd\rangle$ is changed by 60$\%$ of $V_{MU}$. The above
observations suggest that the contribution of $2\hbar\omega$
components is not only A dependent but also state dependent. The
$2\hbar\omega$ components are $4.3\%$, $16.0\%$ and $6.0\%$ in
$3^{+}_{1}$, $1^{+}_{1}$ and $0^{+}_{1}$ states in $^{10}$B,
respectively. It is interesting to do systematic investigation on
how $\langle pp|V|sdsd\rangle$ as well as $2\hbar\omega$ or more
$\hbar\omega$ components affect the energies and effective
operators, such as effective charges and spin g factors which will
be mentioned in the next two sections.

\section{\label{sec:level6}electric quadrupole properties}

\begin{figure*}
\includegraphics[scale=0.8]{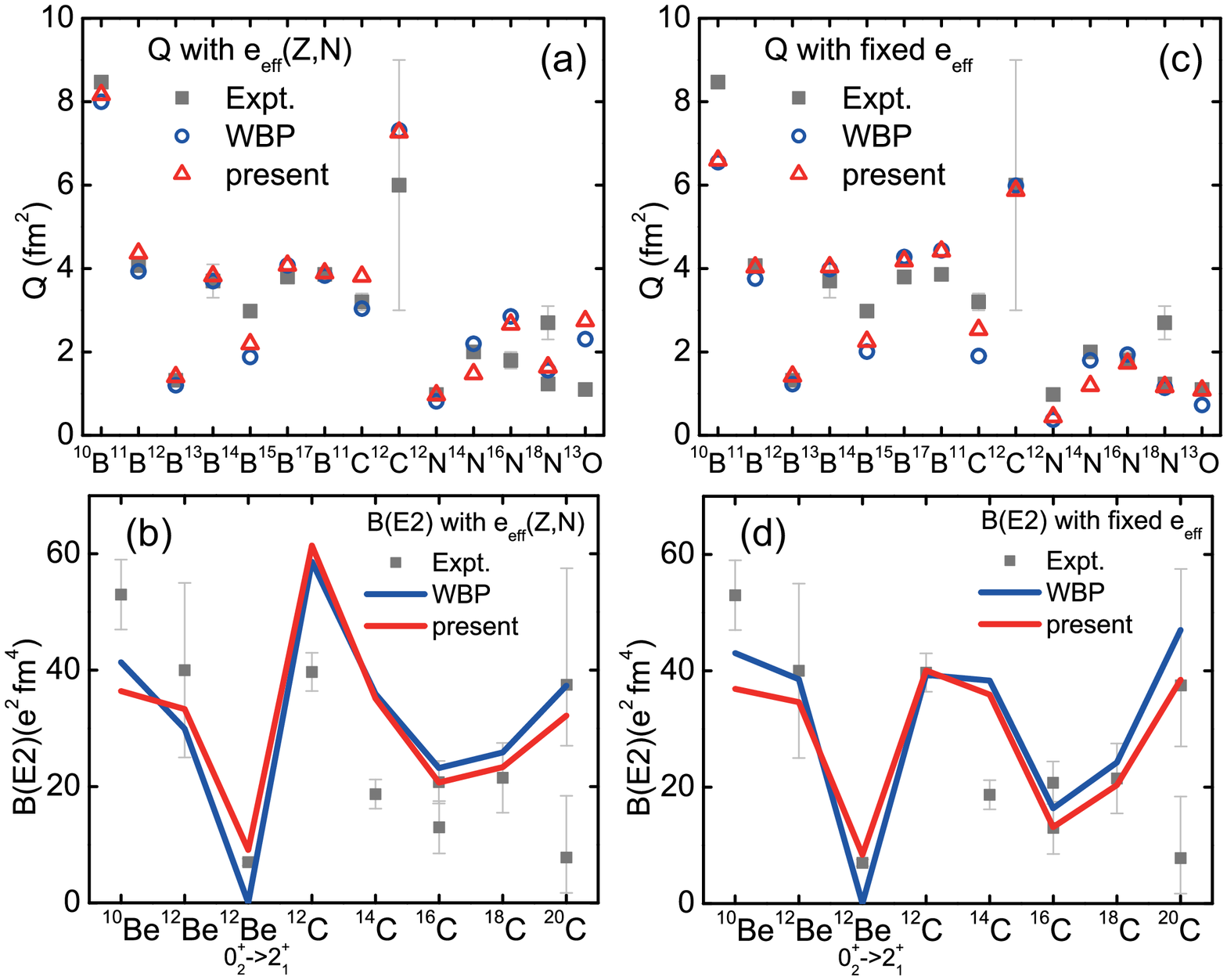}
\caption{\label{E2} (Color online) Electric quadrupole moments Q and
$B(E2)$ values calculated by the present and WBP interactions,
compared with experimental
data~\cite{stone2005,raman2001,shimoura2007,imai2009,wiedeking20082,ong2008,elekes2009,petri2011}.
Two sets of effective charges are used: one is Z, N
dependent~\cite{sagawa2004} and another is fixed to be $e_{p}=1.26$,
$e_{n}=0.21$ and $e_{p}=1.27$, $e_{n}=0.23$ for the present and WBP,
respectively. (a) Electric quadrupole moments calculated with Z, N
dependent effective charges. (b) $B(E2)$ values calculated with Z, N
dependent effective charges. (c) Electric quadrupole moments
calculated with fixed effective charges. (d) $B(E2)$ values
calculated with fixed effective charges. All quadrupole moments are
for the ground states except for $2_{1}^{+}$ in $^{12}$C. All
$B(E2)$ values are from $0_{1}^{+}$ to $2_{1}^{+}$ except for the
second $B(E2)$ value in $^{12}$Be, which is from $0_{2}^{+}$ to
$2_{1}^{+}$. }
\end{figure*}

\begin{figure}
\includegraphics[scale=0.78]{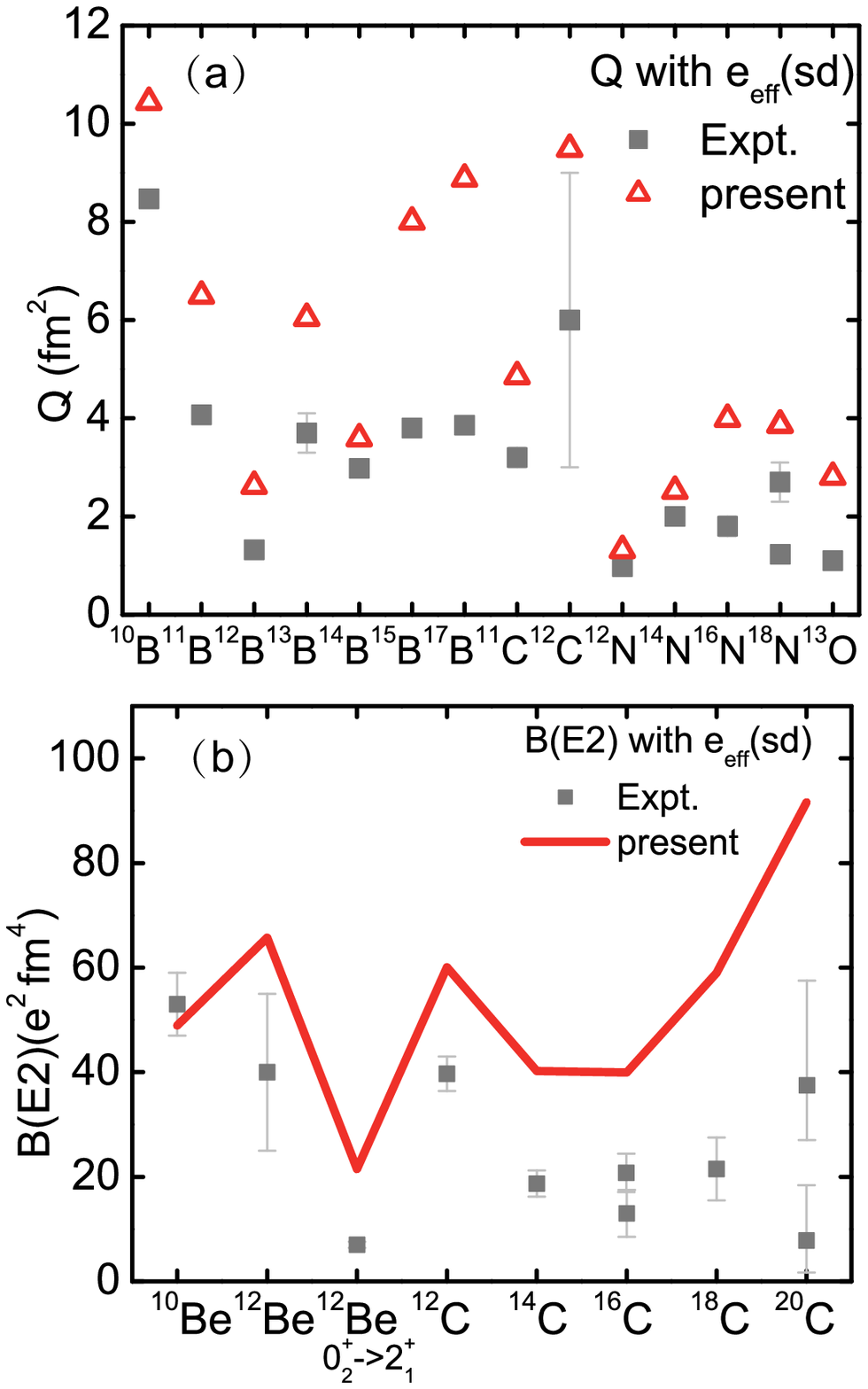}
\caption{\label{E2sde} (Color online) Similar to Fig.~\ref{E2}, but
for $e_{p}=1.3$ and $e_{n}=0.5$.}
\end{figure}

The present Hamiltonian has been shown to be able to describe the
energies of the $psd$-shell nuclei quite well. It is necessary to
investigate whether this interaction gives appropriate wave
functions as well. In this section, we discuss the electric
quadrupole properties with the use of the present Hamiltonian and
WBP. In shell model, effective charges are needed because of the
polarization of the core which is not included in the model
space~\cite{brown2001,Caurier2005}. One set of effective charges,
$e_{p}=1.3$ and $e_{n}=0.5$, is suitable for $sd$-shell
nuclei~\cite{brown2001}, which means that both valence protons and
neutrons are excited in the $sd$-shell. For valence protons and/or
neutrons locate in $p$ shell in neutron-rich nuclei, this set of
effective charges becomes invalid~\cite{sagawa2004,elekes2009}.

Figure~\ref{E2} shows the quadrupole moments in B, C and N isotopes
and $B(E2)$ in Be and C isotopes with two sets of effective charges,
one is Z, N dependent~\cite{sagawa2004} and the other is independent
of Z and N. Experimental values are taken from
Refs.~\cite{stone2005,raman2001,shimoura2007,imai2009,wiedeking20082,ong2008,elekes2009,petri2011}.
For the Z and N independent effective charges, we obtain them by
fitting to quadrupole moments of these nuclei except for $^{18}$N
and $^{10}$B. Quadrupole moment of $^{18}$N is not exactly
determined as there are two experimental values~\cite{stone2005}. In
case of $^{10}$B, Z- or N-independent effective charges cannot
describe well its quadrupole moment, as will be discussed later.

The Z and N independent effective charges obtained for the present
Hamiltonian and WBP are $e_{p}=1.26$, $e_{n}=0.21$ and $e_{p}=1.27$,
$e_{n}=0.23$, respectively. We also get the effective charges for
the present Hamiltonian in $0\hbar\omega$ model space, $e_{p}=1.25$
and $e_{n}=0.25$. The inclusion of the $2\hbar\omega$ model space
reduces the effective charges a little. Both of them underestimate
the quadrupole moments in stable nuclei such as $^{10}$B, $^{11}$C
and $^{12}$N and overestimate those of the nuclei somewhat far from
the stability-line such as $^{15}$B and $^{17}$B. This probably
means that stable nuclei have stronger core polarization while
nuclei far from the stability-line have weaker core polarization. In
nuclei far from the stability-line, some valence nucleons are weakly
bound, which will make the radial wave function extended farther
than the well bound nucleons. The extended wave function will reduce
the interaction between valence nucleons and the core. In case of
$^{12}$C, the results for $B(E2)$ values with fixed effective
charges are better than those with Z and N dependent effective
charges. We emphasize that the smaller but constant effective
charges can reproduce experimental data rather well in
Fig.~\ref{E2}. The smallness may be explained as a consequence of
the small core of $^{4}$He in the present work. More studies on
effective charges are of great interest.

Although none of combinations of WBP or present Hamiltonians  with
either set of effective charges  works well in the quadrupole moment
of $^{14}$B, the present Hamiltonian improves the result of $^{14}$B
compared with WBP. We also calculate this quadrupole moment in
$0\hbar\omega$ model space with the present interaction. The result
becomes worse than that we show in Fig.~\ref{E2} which is obtained
in $2\hbar\omega$ model space. The ground state of $^{14}$B includes
$17\%$ of $2\hbar\omega$ configurations. Including more
$\hbar\omega$ excitations may improve the result of $^{14}$B. The
$2\hbar\omega$ configurations also improve the
$B(E2;0_{2}^{+}\rightarrow2_{1}^{+})$ of $^{12}$Be. The $2_{1}^{+}$
of $^{12}$Be is almost a pure $2\hbar\omega$ state, that is, with
$93\%$ of the $2\hbar\omega$ components in the present Hamiltonian.
The $0_{1}^{+}$ and $0_{2}^{+}$ of $^{12}$Be have $64\%$ and $54\%$
of the $2\hbar\omega$ components, respectively. Therefore, although
$B(E2;0_{1}^{+}\rightarrow2_{1}^{+})$ values are very close to each
other in WBP and the present results, they are contributed by
different configurations in each calculation. In the WBP result,
$B(E2;0_{1}^{+}\rightarrow2_{1}^{+})$ of $^{12}$Be is all from the
contributions by the transition between $p$ shell nucleons,
especially the transition inside $0p_{3/2}$ proton orbit. In the
present result, besides $p$ shell protons, $p$ and $sd$ shell
neutrons contribute a lot to $B(E2;0_{1}^{+}\rightarrow2_{1}^{+})$
in $^{12}$Be. In $^{12}$Be, the pure $p$ shell proton is not enough
to reproduce the $B(E2;0_{2}^{+}\rightarrow2_{1}^{+})$ value as we
see in Fig.~\ref{E2}.

We also try the conventional effective charges for $sd$ shell,
$e_{p}=1.3$ and $e_{n}=0.5$, to calculate quadrupole moments and
$B(E2)$ values with the present Hamiltonian in Fig.~\ref{E2sde}. It
is seen clearly that this set of effective charges is also invalid
for this new Hamiltonian. Almost all values are much overestimated
with this set of effective charge.

\section{\label{sec:level7}spin properties}

\begin{figure}
\includegraphics[scale=0.52]{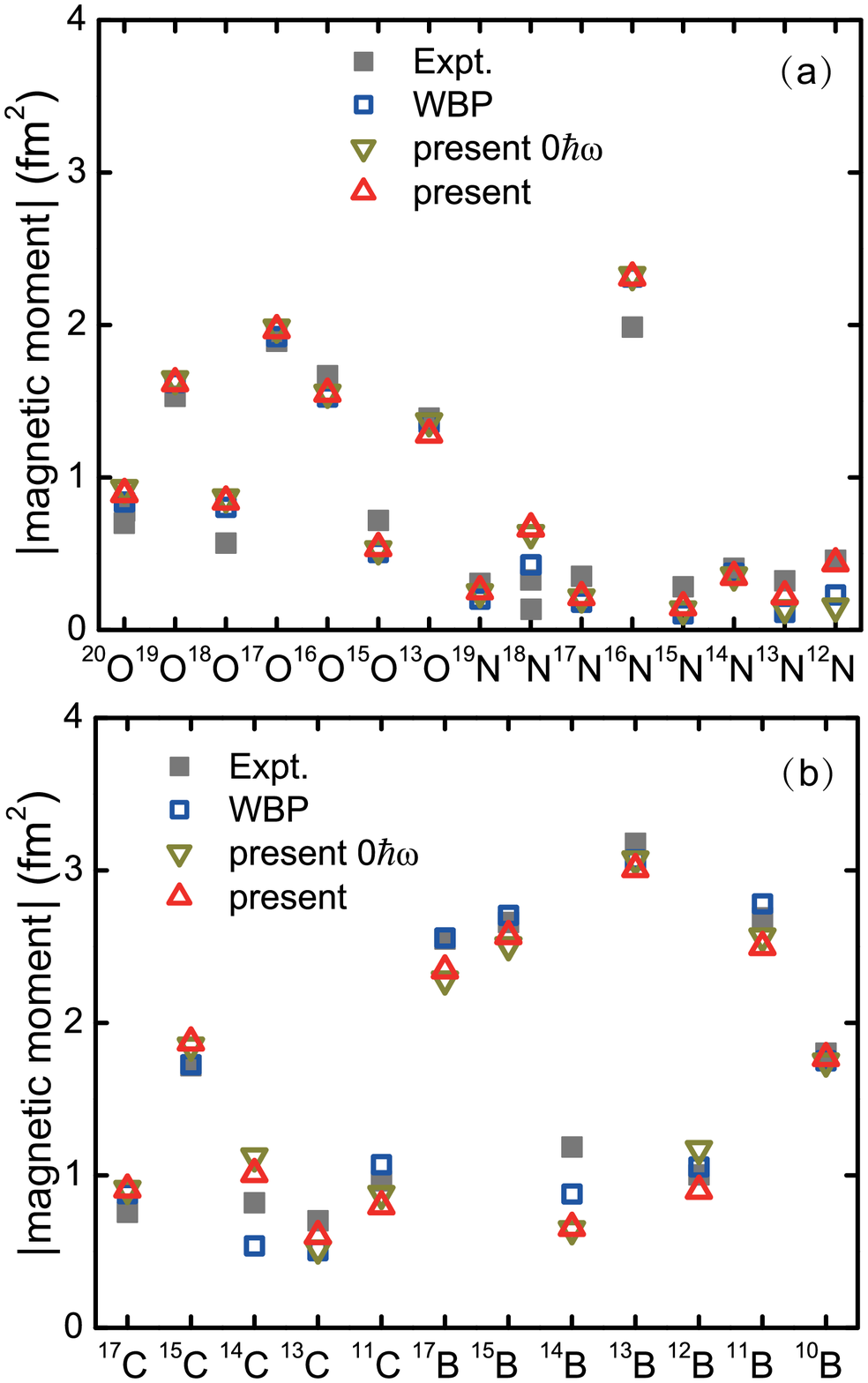}
\caption{\label{m.m.}(Color online) Magnetic moments calculated with
WBP in $0\hbar\omega$, present in both $0\hbar\omega$ and
$2\hbar\omega$ model spaces, compared with experimental
data~\cite{stone2005}. All magnetic moments are for the ground
states except for $2_{1}^{+}$ in $^{20}$O and $^{18}$O, $3_{1}^{-}$
in $^{16}$O and $^{14}$C.}
\end{figure}

\begin{figure}
\includegraphics[scale=0.29]{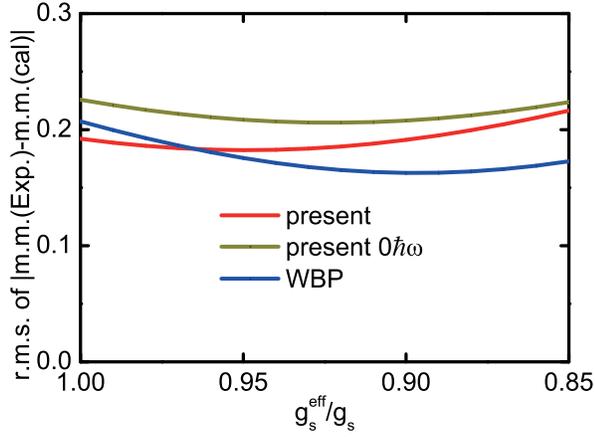}
\caption{\label{m.m.rms} (Color online) Root-mean-square deviation
between calculated magnetic moments (m.m.) and the experimental
ones, $|m.m.(Exp.)-m.m.(Cal.)|$, as the function of
$g^{(eff)}_{s}/g_{s}$.}
\end{figure}

\begin{figure}
\includegraphics[scale=0.36]{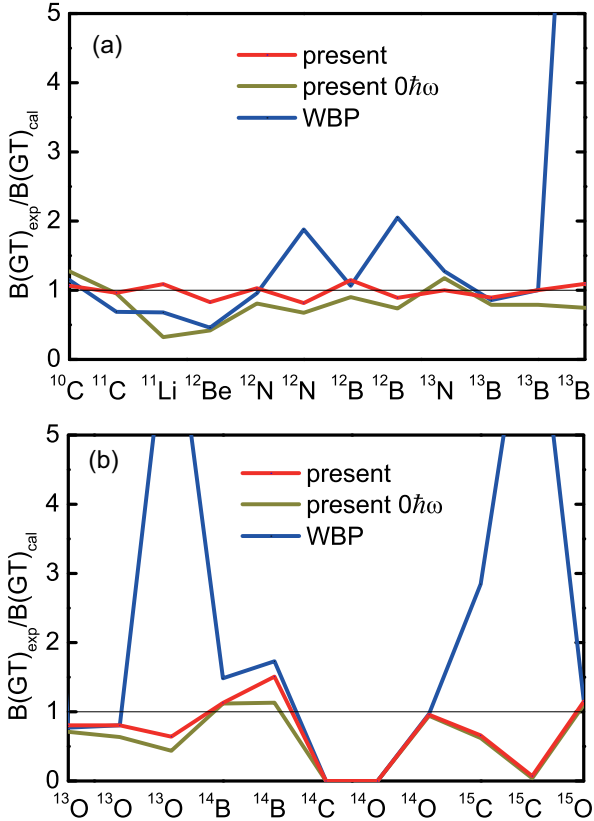}
\caption{\label{BGTg} (Color online) Ratio of the observed $B(GT)$
values over the calculated $B(GT)$, $B(GT)_{exp}/B(GT)_{cal}$, in
nuclei listed in Table~\ref{BGTt}.}
\end{figure}

\begin{figure}
\includegraphics[scale=0.29]{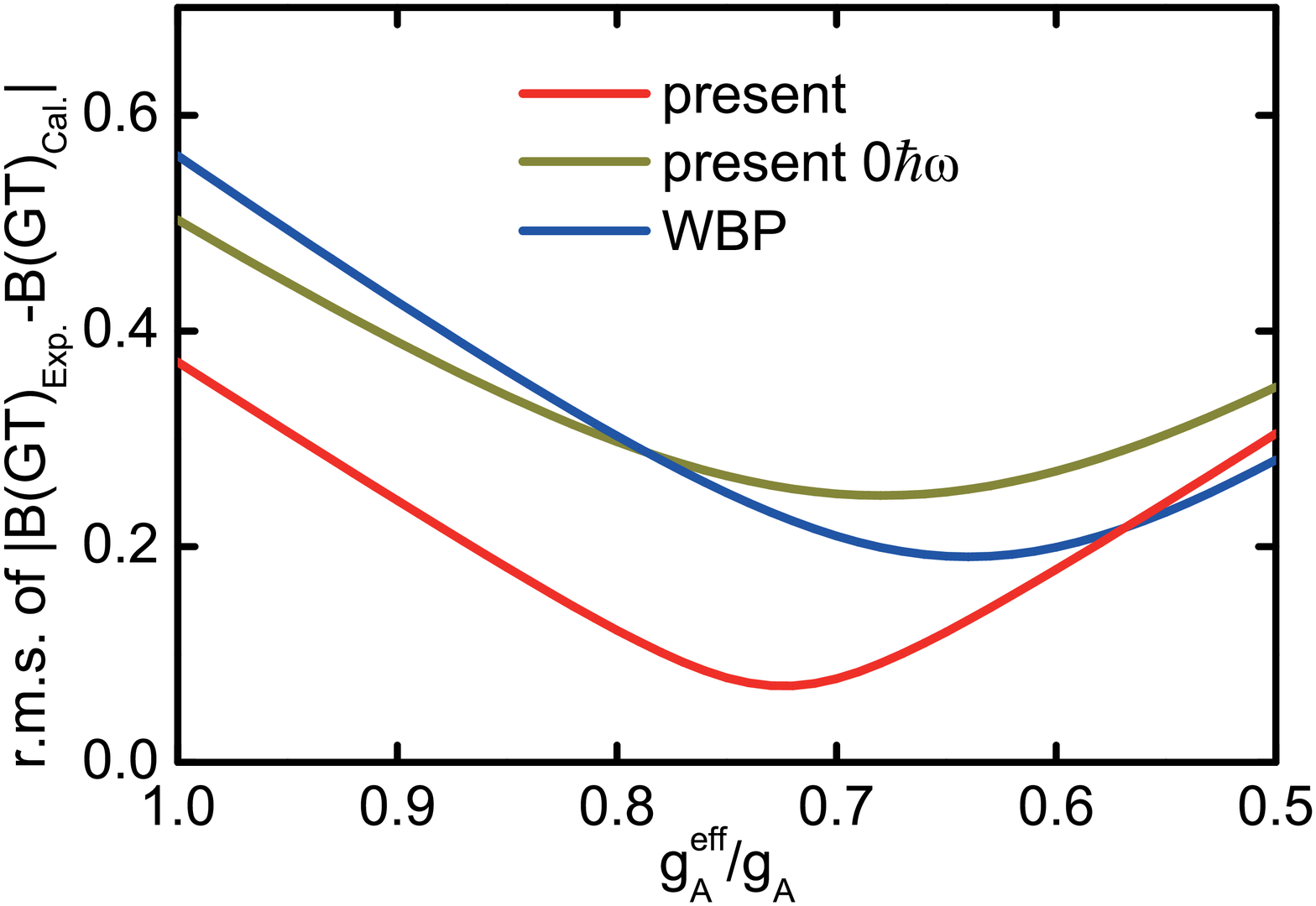}
\caption{\label{rmsBGT} (Color online) Root-mean-square of
$|B(GT)_{cal}- B(GT)_{exp}|$ as the function of
$g^{(eff)}_{A}/g_{A}$.}
\end{figure}

If two protons (neutrons) couple to a pair of angular momentum zero,
their total magnetic moment (m.m.) is zero. The m.m. reflects the
motion of unpaired protons and/or neutrons. Figure~\ref{m.m.}
presents the m.m with WBP and the present Hamiltonian in both
$0\hbar\omega$ and $2\hbar\omega$ model spaces with $\delta
g^{(l)}_{\pi,\nu}=\pm0.1~\mu_{N}$ and $g^{(eff)}_{s}/g_{s}=0.95,
0.92$ and $0.90$ for the present $2\hbar\omega$, the present
$0\hbar\omega$ and WBP, respectively. $\delta g^{(l)}$ comes from
the meson exchange processes~\cite{arima1986,towner1987} and $\delta
g^{(s)}/g^{(s)}$ is obtained from the $\chi$-square fitting of the
calculated values to the experimental ones in these nuclei. All
theoretical results reproduce well the observed values except for a
few nuclei. The largest deviation between calculations and
experimental results is found in $^{18}$O, $^{16}$N, $^{14}$C and
$^{14}$B. Notice that their $E2$ properties are also not well
described. We do not show the result for $^{18}$O in the present
work. Its calculated $B(E2)$ value is much smaller than the observed
one. These nuclei may demand larger model space with $4\hbar\omega$
or more excitations.

\begin{table*}
\caption{\label{BGTt}$B(GT)$ values of experiment, WBP, present in
both $0\hbar\omega$ and $2\hbar\omega$ results. Experimental values
are taken from the Ref.~\cite{suzuki2003} and the related references
in this paper.}
\end{table*}

In Fig.~\ref{m.m.rms}, we also show the root-mean-square (r.m.s.)
deviation between calculations and observed values,
$|m.m.(Exp.)-m.m.(Cal.)|$, versus $g^{(eff)}_{s}/g_{s}$. In a region
$\delta g^{(eff)}_{s}/g_{s}=\pm0.03$, the r.m.s. deviation of each
result is very flat. Outside this region, the r.m.s. deviation
increases. The minimal point for the r.m.s. deviation is located at
$g^{(eff)}_{s}/g_{s}=0.95, 0.92$ and $0.90$ for the present
$2\hbar\omega$, the present $0\hbar\omega$ and WBP, respectively. As
we expect, the quenching is weaker when we enlarge the model space.
The quenching of the present $2\hbar\omega$ result is rather weak
and we may safely use bare $g_{s}$. If all results are with bare
$g_{s}$, the present Hamiltonian gives the smallest r.m.s.
deviation. We should also note, on the other hand, that the
quenching factor obtained here has some ambiguity as the dependence
of the r.m.s. deviation on the value of $g_s^{eff}/g_s$ is quite
modest.

Table~\ref{BGTt} presents the Gamow-Teller transition rates $B(GT)$.
The $B(GT)$ values can be extracted from experimental $log ft$ values with
the equation,
\begin{equation}\label{coulomb}
ft=\frac{6147}{(g_{A}/g_{V})^{2}B(GT)},
\end{equation}
where $6147$ is from Ref.~\cite{hardy1990}, $g_{A}$ and $g_{V}$ are
the axial-vector and vector coupling constants, respectively. For
beta decays, we use bare $g_{A}/g_{V}=-1.26$~\cite{bopp1986}. The
calculated results are with $g_{A}^{eff}$ which is from
$\chi$-square fitting of these $B(GT)$ values. The
$(g_{A}^{eff}/g_{A})=0.72, 0.68$ and $0.64$ for the present
$2\hbar\omega$, the present $0\hbar\omega$ and WBP, respectively.
The $(g_{A}^{eff}/g_{A})$ value for WBP is very close to the
commonly used value $0.60$~\cite{brown2001}. $^{11}$Li and $^{15}$C
are weakly bound with $0.325$ and $1.218$ MeV neutron separation
energy, respectively. The protons in their daughter nuclei from
$\beta$ decay are well bound. Halo or skin effects need to be
included which is not included in calculations with harmonic
oscillator bases. The overlap between related neutron and proton
orbits is calculated in Woods-Saxon bases to modify the $B(GT)$ of
these two nuclei. All these three calculated results, WBP, the
present in $2\hbar\omega$ and $0\hbar\omega$, are modified by the
halo or skin. More details can be found in Ref.~\cite{suzuki2003}.

The present $2\hbar\omega$ results improve most of the $B(GT)$
values compared with the present $0\hbar\omega$ and WBP results. In
order to show the difference between calculations and observed
values, we present $B(GT)_{exp}/B(GT)_{cal}$ in Fig.~\ref{BGTg}.
$B(GT)_{exp}/B(GT)_{cal}$ from the present $2\hbar\omega$
calculation is very close to unity except for $^{14}$C and $^{14}$O,
and the second transitions in $^{15}$C. $^{14}$C and $^{14}$O are
the same in the present isospin symmetric Hamiltonian. The
abnormally long lifetime of $^{14}$C has been a long-standing
theoretical problem~\cite{talmi2003}. The present $0\hbar\omega$ and
WBP results also fail in describing $B(GT)$ of $^{14}$C and
$^{14}$O. The reason is that two main components of the transition
are almost all canceled in $^{14}$C~\cite{talmi2003}. It is hard to
describe the cancellation exactly in interactions determined by
considering all nuclei nearby. In case of the second transition from
$^{15}$C to $^{15}$N, the reason is similar, that is, three
components are canceled resulting in a rather small value.

Similar to the discussion in m.m., the r.m.s. deviation of
calculated $B(GT)$ values from the experimental ones is presented in
Fig.~\ref{rmsBGT}. It is clearly seen that both the
 r.m.s. deviation and the quenching get smaller when the model space is enlarged.

\section{\label{sec:level8}Summary}
In the present work, we present a systematic study of boron, carbon,
nitrogen and oxygen nuclei in full $psd$ model-space with a newly
constructed Hamiltonian. While some former Hamiltonians, such as
PSDMK, WBP and WBT, are constructed in $0-1\hbar\omega$ model space,
we include $2-3\hbar\omega$ excitations in the present work. The
present Hamiltonian is based on $V_{MU}$ and have four parts,
$\langle pp|V|pp\rangle$ from SFO, $\langle sdsd|V|sdsd\rangle$ from
SDPF-M, $\langle psd|V|psd\rangle$ and $\langle pp|V|sdsd\rangle$
from $V_{MU}$ plus spin-orbit force. We optimize the central part of
$V_{MU}$ while the tensor force in $V_{MU}$ and the spin-orbit force
are kept unchanged. The central force in $\langle psd|V|psd\rangle$
is $30\%$ of $V_{MU}$'s stronger than that in $\langle
pp|V|sdsd\rangle$, while the strength of these two parts are the
same in WBP. The SPE of the five orbits are also modified. More
details of this Hamiltonian are explained in the text.

The present Hamiltonian can reproduce well the ground-state
energies, drip lines, energy levels, electric properties and spin
properties of $psd$-shell nuclei. Especially, we can describe the
drip lines of carbon and oxygen isotopes and spins of the ground
states of $^{10}$B and $^{18}$N where WBP and WBT fail. The
inclusion of $2\hbar\omega$ excitations is important in describing
such properties because a part of mass-dependent effect in WBP and
WBT is naturally included when we include $2\hbar\omega$
excitations. The effective operators become closer, in general, to
bare operators when we enlarge the model space. We note that
constant and smaller effective charges work quite well in the
present study, which may attributed to the small size of the
$^{4}$He core also. The contribution coming from $2\hbar\omega$
excitations are investigated by comparison to $0\hbar\omega$
calculations, suggesting that the present model space is still
insufficient to reduce effective charges almost to zero. More
systematic study is needed in a model space larger than $psd$ and
more $\hbar\omega$ excitations, especially for $4p4h$ excitations
from $p$ to $sd$ shells.

It is also examined whether tensor force and spin-orbit force can be
kept unchanged in full shell-model calculations. Shell-model
calculations without the modification of the strength of these two
forces are found to be successful in the description of a wide range
of $psd$-shell nuclei. It is interesting to do more work on applying
the present tensor and spin-orbit forces to shell-model calculations
in other region of nuclei.

\section*{Acknowledgement}
The shell-model calculations in this work are made by the codes
OXBASH~\cite{OXBASH}. This work has been supported by the National
Natural Science Foundation of China under Grant No. 10975006. It has
also been supported in part by Grants-in- Aid for Scientific
Research (A) 20244022 and (C) 22540290 of the Ministry of Education,
Culture, Sports, Science and Technology of Japan. The author C. Y.
thanks for the financial support from China Scholarship Council.

\end{document}